 \documentclass[final,5p,times,twocolumn]{elsarticle}

 \usepackage{graphicx}

\usepackage{amssymb}
\usepackage{amsmath}

\biboptions{sort&compress}

\journal{Chem. Phys. Lett.}

\renewcommand{\vec}[1]{\mathbf{#1}}      
\allowdisplaybreaks
\let\leq\leqslant

\begin{document}

\begin{frontmatter}



\title{Structural phase transition and band gap of uniaxially deformed $(6,0)$ carbon nanotube}


\author[BSU]{Nikolai A. Poklonski\corref{cor}}
\ead{poklonski@bsu.by}
\author[BSU]{Sergey V. Ratkevich}
\author[BSU]{Sergey A. Vyrko}
\author[BSU]{Eugene~F.~Kislyakov}
\author[BSU]{Oleg N. Bubel'}
\address[BSU]{Physics Department, Belarusian State University, 220030 Minsk, Belarus}
\author[ISAN]{Andrei M. Popov}
\ead{am-popov@isan.troitsk.ru}
\address[ISAN]{Institute of Spectroscopy, 142190 Troitsk, Moscow Region, Russia}
\author[ISAN,MIPT]{Yurii~E.~Lozovik}
\address[MIPT]{Moscow Institute of Physics and Technology, 141701, Dolgoprudny, Moscow Region, Russia}
\author[SGU]{Nguyen Ngoc Hieu}
\address[SGU]{Physics Department, Sai Gon University, Ho Chi Minh City, Vietnam}
\author[IPE]{Nguyen Ai Viet}
\address[IPE]{Institute of Physics and Electronics, Hanoi, Vietnam}
\cortext[cor]{Corresponding author}

\begin{abstract}
The atomic and band structures of the $(6,0)$ \emph{zigzag} carbon nanotube at its axial elongation are calculated by semiempirical molecular orbital and by tight-binding methods. The ground state of the nanotube is found to have a Kekule structure with four types of bonds and difference between lengths of long and short bonds of about 0.005~nm. The structural phase transition is revealed at ${\approx}\,9$\% elongation, resulting in a quinoid structure with two types of bonds. This structural phase transition is followed by the transition from a narrow gap to moderate gap semiconductor. Validity of the semiempirical PM3 method is discussed.
\end{abstract}

\begin{keyword}
Zigzag carbon nanotube \sep Molecular orbital calculations \sep Kekule structure \sep Phase transition \sep Tight-binding approximation \sep Band gap

\end{keyword}

\end{frontmatter}

\hyphenation{nano-tube nano-tubes}


\section{Introduction}
The studies of structural, electronic and elastic properties of carbon nanotubes (CNTs) are actual in connection with perspectives of their applications in nanoelectronic devices and in composite materials~\cite{Wong11}. These properties are also of fundamental interest, particularly for physics of phase transitions. For example, such structural phase transitions as the commensurate\hspace{0pt}--\hspace{0pt}incommensurate phase transition in doublewalled CNTs~\cite{Bichoutskaia06, Popov09} and the spontaneous symmetry breaking with formation of corrugations along nanotube axis~\cite{Connetable05} have been considered.

The possibility of Peierls transition in CNTs was first considered in~\cite{Mintmire92}. As a result of this transition, metallic CNTs become semiconducting at low temperature and Peierls distortions of the nanotube lattice lead to a Kekule structure (see Fig.~1a). The Kekule structure can arise in a $(m,n)$ CNT in the case that $m - n$ is multiples of 3~\cite{Harigaya93}. The Peierls gap~\cite{Harigaya93, Viet94, Huang96}, Peierls transition temperature~\cite{Mintmire92, Harigaya93, Huang96, Sedeki00, Bohnen04, Figge01, Chen08} and bond lengths changes~\cite{Harigaya93, Viet94, Poklonski08CPL187} due to Peierls distortions were estimated for $(n,n)$ \emph{armchair} CNTs. Static twist deformation of the $(5,5)$ \emph{armchair} nanotube lattice at the Peierls transition was also predicted~\cite{Figge01, Chen08}. The other type of Peierls distortions was found for the semiconductor chiral CNTs~\cite{Harigaya93, Tretiak07}. The Kekule structure of the ground state was found for the finite length \emph{armchair} CNTs using density functional theory~\cite{Zhou04, Matsuo03, Nakamura03} and semiempirical molecular orbital~\cite{Nakamura03} calculations. X-ray crystallographic analysis of chemically synthesized short $(5,5)$ CNTs shows the Kekule bond length alternation pattern for their structure~\cite{Nakamura03}.

Although the Kekule structure can arise in $(n,0)$ \emph{zigzag} CNTs, if $n$ is multiples of 3, such possibility has not yet been considered. The density functional theory calculations performed for short \emph{zigzag} CNTs show that pattern of short and long bonds arise~\cite{Yumura04, Yumura04JPC11426}, however this pattern cannot be assigned to the Kekule structure (as it is defined in~\cite{Harigaya93, Okahara94}). Previously~\cite{Poklonski08CPL187} we have revealed the Kekule structure of the ground state of $(5,5)$ \emph{armchair} CNT by semiempirical molecular orbital calculations and found two deformational structural phase transitions connected with spontaneous symmetry breaking and controlled by uniaxial deformation of the CNT. 

In the present Letter we consider the possibility of the Kekule structure of the ground state and structural phase transitions in $(3n,0)$ \emph{zigzag} CNTs on the example of the infinite $(6,0)$ CNT. We present the semiempirical molecular orbital PM3~\cite{Stewart89} calculations of the internal energy and atomic structure of a uniaxially deformed infinite $(6,0)$ CNT up to 12\% elongation. For calculated atomic structure we use simple tight-binding (H\"uckel)~\cite{Saito98} calculations to gain qualitative insight into changes of the electron energy band structure of a uniaxially deformed $(6,0)$ CNT at structural phase transitions. Geometry optimization of an infinite $(6,0)$ CNT was also performed by density functional theory (DFT) calculations.

\section{Methodology}

Semiempirical molecular orbital method PM3~\cite{Stewart89} modified for one-dimensional infinite periodic structures~\cite{Stewart87} is used to calculate the internal energy and atomic structure of the $(6,0)$ CNT under uniaxial strain. Calculations were performed using the MOPAC2007 code~\cite{MOPAC2007}.

The adequacy of the PM3 parameterization of the Hamiltonian has been already demonstrated~\cite{Bubel00} by the calculation of bond lengths of the C$_{60}$ fullerene with $I_h$ symmetry: the calculated values of the bond lengths coincide with the measured ones~\cite{Leclercq93} at the level of experimental accuracy of $10^{-4}$~nm at liquid helium temperature. The PM3 parameterization gives also correct bond length in graphite~\cite{Budyka05}. Comparison of different types of semiempirical calculations~\cite{Stewart07} shows that the PM3 parameterization gives the best accuracy for structure and internal energy of carbon nanostructures.

To check the convergence of calculated quantities versus the size of the computational cell the structure of the ground state of the infinite undeformed $(6,0)$ CNT is obtained with the use of computational cells comprising 72, 96 and 120 atoms corresponding to three, four and five translational periods of the CNT, respectively. The results presented below are obtained for the computational cell comprising 96 atoms that allows to calculate the internal energy and bond lengths with the accuracy within 0.4\% and 0.1\%, respectively. Calculations were performed for the infinite CNT with the use of Born--von Karman boundary conditions along the nanotube axis with full geometry optimization (without any symmetry constraints) including the computational cell length.

For the computational cell comprising 96 carbon atoms, we have also performed DFT calculations of the nanotube geometry within package Quantum-ESPRESSO (program PWscf for plane wave pseudopotential method)~\cite{Giannozzi09}. Periodic boundary conditions along the nanotube axis with full geometry optimization were used, spin-polarization was not taken into account.

PWscf program was configured to use Perdew--Zunger \cite{Perdew81} local density approximation (LDA) of exchange-correlation energy (pseudopotential C.pz-vbc.UPF). Energy cutoff of plane waves was set to 408~eV. Geometry optimization was performed within a quasi-Newton scheme combined with the Broyden--Fletcher--Goldfarb--Shanno algorithm for Hessian updating (see also~\cite{Burke12}). We have also checked DFT calculations by use of Perdew--Burke--\hspace{0pt}Ernzerhof (PBE) parametrization~\cite{Perdew96} of the generalized gradient approximation (GGA) of exchange-correlation energy. It was found no significant difference between the results of GGA and LDA calculations.

In order to check the accuracy of the used DFT method we have calculated geometry of fullerene C$_{60}$ and compared obtained bond lengths with the experiment~\cite{Leclercq93}. Geometry optimization by PWscf showed certain dimerization of C--C bond lengths in C$_{60}$ molecule. The deviation of the PWscf calculations from the average measured values was within 1.3\%, while for PM3 it was less than 0.2\%.

\section{Ground state structure of $(6,0)$ nanotube}

Contrary to the \emph{armchair} CNTs~\cite{Harigaya93, Poklonski08CPL187}, the multiplication of the translational period along the nanotube axis is not necessary for \emph{zigzag} CNTs with the Kekule structure. In the work~\cite{Torres06} a possibility of Peierls gap opening was considered in the electron energy spectrum of the $(3n,0)$ CNTs in the dynamic regime (due to the interaction of electrons with longitudinal optical phonons at the boundary of the Brillouin zone). Since the wavelength of such phonons is twice as large as a translational period of the CNT, this leads to the doubling of the translational period of the system. In order to check whether the static Peierls distortions are realized for the $(6,0)$ CNT, we performed PM3 calculations with the length of computational cell equal to four translational periods of the CNT. These calculations as well as the calculations with the length of computational cell equal to three and five periods reveal ordinary translational period along nanotube axis for the $(6,0)$ CNT.

Our calculations give the Kekule structure for the $(6,0)$ CNT at the ground state ($\varepsilon = 0$), see Fig.~1a. There are four types of bonds which are arranged alternatively in the lattice. The bonds $a$ and $c$ are directed along and the bonds $b$ and $d$ are skewed with respect to the nanotube axis. The considerable difference of 0.005~nm approximately is found between lengths of short bonds $a \approx b$ and long bonds $c \approx d$.

Contrary to the PM3 method, the DFT calculations show only the quinoid structure of the $(6,0)$ CNT and do not show the Kekule structure for any elongation of the nanotube. However, the DFT method seems to be less accurate in geometry calculations of carbon nanostructures than the PM3 method as being compared to experimental data (see the last paragraph of the methodology section).

\begin{figure}
\hfil\includegraphics{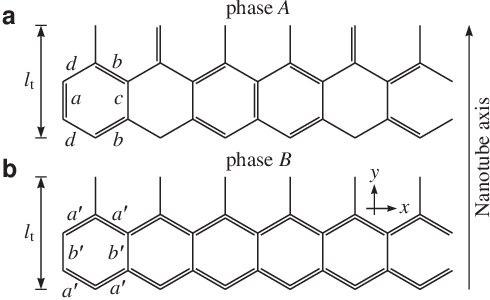}
\caption{The translational unit cell of the (6,0) nanotube, unfolded on the figure plane, at the ground state without deformation (a) and at relative axial elongation $\varepsilon = 10\%$ (b); $l_{\rm t}$ is the translational period along the nanotube axis $y$. Bond lengths: (a) Kekule structure: $a = 0.1393$~nm, $b = 0.1408$~nm and $c = 0.1433$~nm, $d = 0.1447$~nm; (b) quinoid structure: $a' = 0.1449$~nm and $b' = 0.1541$~nm.}\label{fig:01}
\end{figure}

\section{Deformational structural transition in $(6,0)$ nanotube}

\begin{figure}
\hfil\includegraphics{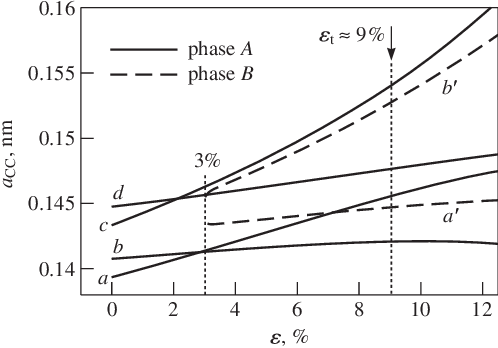}
\vspace{-6pt}
\caption{Dependence of the bond lengths $a$, $b$, $c$, $d$, $a'$ and $b'$ of the $(6,0)$ nanotube on the relative axial elongation $\varepsilon$. Phase $A$ corresponds to the Kekule structure, phase $B$ corresponds to the quinoid structure. Left vertical dotted line indicates the lost of stability for the metastable phase $B$ at the relative elongation $\varepsilon \approx 3\%$. Right vertical dotted line indicates the phase transition at the critical relative elongation $\varepsilon_{\rm t} \approx 9\%$.}\label{fig:02}
\end{figure}

The structure and internal energy of the $(6,0)$ CNT are calculated with full geometry optimization for relative axial elongations of the CNT $\varepsilon ={}$0--12\% ($\varepsilon = (l - l_0)/l_0$, where $l$ is the length of the computational cell and $l_0 = 1.703$~nm is the length of the computational cell at the ground state). The computational cell comprising $N = 96$ carbon atoms with the length of four translational periods of CNT along the nanotube axis is used.

The high kinetic barrier for formation of Stone--Wales defects excludes their spontaneous formation at room temperature up to the elongation $\varepsilon = 15\%$~\cite{Samsonidze02} (see also consideration of these defects formation at bending deformation of CNT~\cite{Moliver11}). Thus, we do not consider nanotube structures with Stone--Wales defects. Note also that we consider CNTs at elongations which are well under the fracture value ($\varepsilon = 16\%$ according to AM1 calculations~\cite{Dumitrica03} and $\varepsilon = 20\%$ according to PM3 and density functional theory-based calculations~\cite{Mielke04}).

The following phases with different symmetries of the structure have been found for the uniaxially deformed $(6,0)$ CNT: the phase $A$ with the Kekule structure at all considered elongations $\varepsilon$ from 0\% to 12\% and the phase $B$ with the quinoid structure at the elongations $\varepsilon$ from 3\% to 12\%. The quinoid structure of the CNT at the great elongation is shown in Fig.~1b. The PM3 calculated dependences of the bond lengths on the elongation $\varepsilon$ are presented in Fig.~2. 

The PM3 calculations of the nanotube internal energy $U$ of the $(6,0)$ CNT at zero temperature as a function of square of the relative elongation is shown in Fig.~3a. Since we study the system at constant length and temperature, the ground state is determined by the minimum of the system internal energy. The difference of the internal energies $U_A - U_B$ of the phases $A$ and $B$ is shown in Fig.~3b. The change of sign of the energy difference $U_A - U_B$ means that the structural phase transition takes place at zero temperature at the critical elongation $\varepsilon_{\rm t} \approx 9\%$. Thus, at zero temperature, the phase $A$ is stable at the elongation $\varepsilon < \varepsilon_{\rm t}$ and the phase $B$ (see Fig.~1b) is stable at $\varepsilon > \varepsilon_{\rm t}$. The phase $A$ at elongation $\varepsilon > \varepsilon_{\rm t}$ and the phase $B$ at elongation $\varepsilon < \varepsilon_{\rm t}$ are metastable.

\begin{figure}[!t]
\hfil\includegraphics{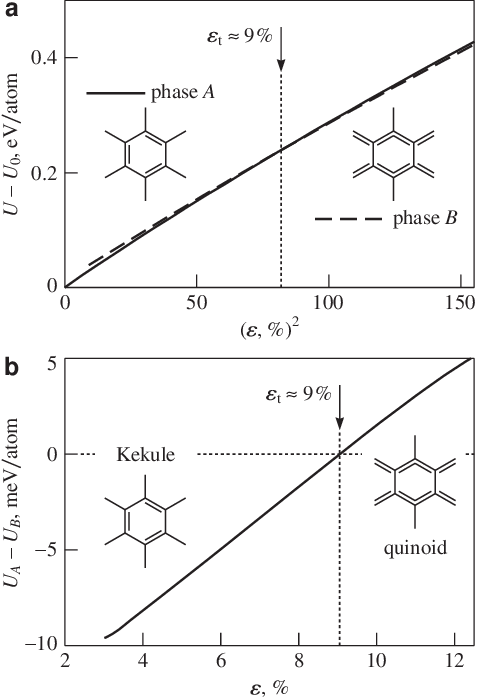}
\caption{Dependence of the internal energy $U$ of the $(6,0)$ nanotube on the square of the relative elongation $\varepsilon^2$ at zero temperature (a) and dependence of the difference of the internal energies of the phases $A$ and $B$ on the elongation $\varepsilon$ (b). Vertical dotted line indicates the phase transition at the critical relative elongation $\varepsilon_{\rm t} \approx 9\%$.}\label{fig:03}
\end{figure}

The dependence of the bond lengths (Fig.~2) on the elongation $\varepsilon$ demonstrates that the transition from the Kekule structure with four types of bonds to the quinoid structure with two types of bonds takes place with abrupt change of the structure of the CNT. Thus, first-order structural phase transition controlled by the nanotube elongation takes place, at zero temperature at the critical elongation $\varepsilon_{\rm t} \approx 9\%$ where the internal energies of the phases coincide (see Fig.~3). At nonzero temperature, phase transitions in one-dimensional systems have a crossover character (see discussion for structural phase transitions in \emph{armchair} CNTs in~\cite{Poklonski08CPL187} and references therein). The elastic constants (coefficient of elasticity, Poisson ratio, torsion modulus) and specific heat of the CNT also abruptly change at the phase transition. Thus, the structural phase transition can be determined by peculiarities of these quantities on temperature or the elongation.

The dependence of the internal energy $U$ on the length $l = (1 + \varepsilon)l_0$ of the computational cell is interpolated by the Hooke's law for the elongation range $\varepsilon = 0$--1\%
\begin{equation}\label{eq:01}
   U - U_0 = \frac{\nu(l - l_0)^2}{2N} = \frac{\nu\varepsilon^2l_0^2}{2N},
\end{equation}
where $\nu$ is the coefficient of elasticity of a CNT with the length $l_0 = 1.703$~nm at the ground state and number of atoms $N$.

\begin{figure*}
\hfil\includegraphics{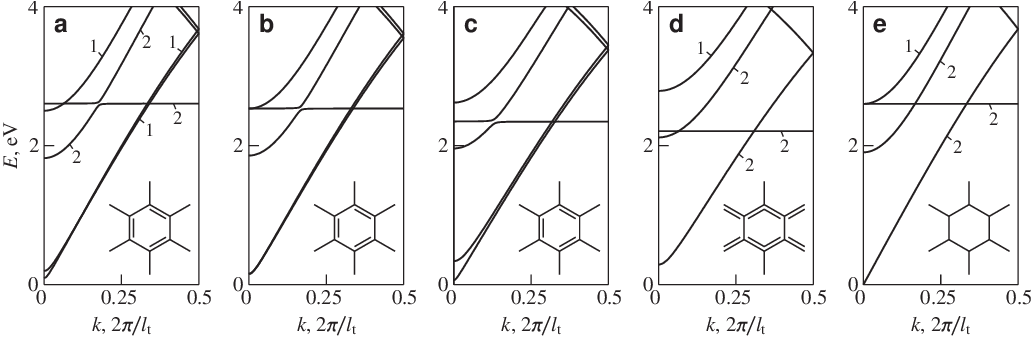}
\caption{The band structure of the (6,0) nanotube with the Kekule structure (a--c), the quinoid structure (d), and the structure with all equal bond lengths (e) at different elongations: (a) $\varepsilon = 0$ (ground state), (b) $\varepsilon = 2.3\%$, (c) $\varepsilon = 8\%$, (d) $\varepsilon = 10\%$, (e) $\varepsilon = 0$. The numbers in parts (a), (d) and (e) indicate degeneracy of energy bands. Only conduction bands are shown. The valence bands are the same as the conduction bands, but with opposite sign. Insets show corresponding bond alternation patterns.}\label{fig:04}
\end{figure*}

Taking into account the coefficient of elasticity $\nu$ defined by Eq.~(\ref{eq:01}), the Young's modulus of a CNT at the elongation $\varepsilon = 0$ has the following form:
\begin{equation}\label{eq:02}
   Y = \frac{\nu l_0}{2\pi Rw},
\end{equation}
where $R = 0.241$~nm is the radius of the CNT corresponding to the elongation $\varepsilon = 0$ and $w = 0.34$~nm is the effective thickness of the wall~\cite{Mielke04, Meo06, Ranjbartoreh10, Krishnan98, Treacy00}. (In Eq.~(\ref{eq:02}) the quantity $2\pi Rw$ is the cross-section area of a nanotube.) As a result, the value $Y \approx 1.2$~TPa is calculated. The obtained value of Young's modulus of the $(6,0)$ CNT with the Kekule structure is in good agreement with the values for this CNT with the quinoid structure: 0.9~TPa by finite element modelling based on molecular mechanics~\cite{Meo06}, 1.42~TPa by molecular dynamics simulation using LAMMPS~\cite{Ranjbartoreh10} and 0.95~TPa by our PWscf DFT calculations. In addition, it agrees well with experimentally determined value 1.25~TPa for singlewalled CNTs of diameters about 1~nm~\cite{Krishnan98, Treacy00}. Note, that this result correlates also with calculations for another \emph{zigzag} nanotubes: 1.1~TPa for the $(10,0)$ CNT by PM3 calculations~\cite{Mielke04}, 1.14~TPa for the $(9,0)$ CNT by Hartree--Fock method~\cite{VanLier00}, 1.05~TPa for the $(10,0)$ CNT~\cite{SanchezPortal99}, 0.94~TPa for the $(10,0)$ CNT~\cite{Mielke04} and 1.04~TPa for the $(9,0)$ CNT~\cite{Bichoutskaia06PRB045435} by density functional theory-based calculations.

Since the radius $R$ depends on the elongation $\varepsilon$, it is not convenient to use Eq.~(\ref{eq:02}) for strained CNTs. Moreover, the coefficient of elasticity $\nu$ is the quantity which can be measured experimentally. Thus, we have calculated the abrupt change of the coefficient of elasticity at the structural phase transition: $\nu_B = 0.97\nu_A$, where $\nu_A$ and $\nu_B$ are the coefficients of elasticity for the phases $A$ and $B$, respectively (the coefficients of elasticity are calculated for the elongation range $\varepsilon = 8$--10\%).

\section{Electronic structure of deformed $(6,0)$ CNT}

Both \emph{zigzag} and \emph{armchair} CNTs with the Kekule structure are semiconducting at the ground state with the band gap being proportional to the difference of the hopping (resonance) integrals corresponding to bonds of different types~\cite{Okahara94}, while the same CNTs with all bonds of equal length are metallic~\cite{Okahara94}. (Note, that tight-binding calculations, which take into consideration not only three nearest atoms but also two next-nearest ones~\cite{Ribeiro10}, also give zero band gap for $(6,0)$ \emph{zigzag} CNT with all equal bond lengths.) Thus, the accurate calculations of the atomic structure of the deformed CNT are necessary to calculate the band structure. To gain a qualitative insight in the energy spectrum of $\pi$-electrons in the uniaxially deformed $(6,0)$ CNT, we perform the tight-binding calculations~\cite{Saito98} for the PM3 calculated atomic structure. Only interaction between nearest-neighbour atoms is taken into account. The hopping integral is used in the Harrison's form~\cite{Harrison89} $t = t_0(a_0/a_{\rm CC})^2$, where $t_0 = 2.6$~eV is the standard value of the hopping integral in a CNT for the C--C bond length $a_0 = 0.142$~nm~\cite{Wildoer98, Odom98}, and $a_{\rm CC}$ is the calculated bond length.

The band structure of the deformed $(6,0)$ CNT with the bond lengths and angles taken from the PM3 calculations is obtained by the diagonalization of the matrixes (\ref{eq:04}) and (\ref{eq:05}) for $\varepsilon = 0$--9\% and $\varepsilon > 9\%$, respectively, and shown in Fig.~4. It is revealed that the deformed $(6,0)$ CNT is semiconducting for all considered elongations.

All $(3n,0)$ \emph{zigzag} CNTs with equal bond lengths were found to be metallic at zero elongation~\cite{Saito98} (the calculated band structure of the $(6,0)$ CNT with all bonds of equal length $a_{\rm CC} = 0.142$~nm is shown in Fig.~4e). Here we have performed the tight-binding calculations of the band structure of the uniaxially deformed $(6,0)$ CNT with equal bond lengths (that is the atomic structure changes only due to the changes of angles between bonds,
\[
   \gamma = 2\arcsin[(3(\varepsilon + 1))^{1/2}/2]),
\]
while bond lengths remain equal. It is found that the band structure of such CNT does not change with the elongation.

The calculated band gap (in the center of the Brillouin zone) in the electron energy spectrum of the $(6,0)$ CNT as a function of the elongation $\varepsilon$ is shown in Fig.~5. For the Kekule structure the band gap linearly increases (from 187~meV to 311~meV) with the elongation for the elongation range $\varepsilon = 0$--2.3\% and linearly decreases (from 311~meV to 88~meV) for the elongation range $\varepsilon = 2.3$--9\%. The bond lengths linearly increase with the elongation. The phase transition from the narrow band gap ($\approx 0.1$~eV) to the moderate band gap ($\approx 0.5$~eV) semiconductor occurs in the electron subsystem of the CNT at the elongation $\varepsilon_{\rm t} \approx 9\%$ corresponding to the structural phase transition. Unlike the Kekule structure, the band gap of the quinoid structure increases (from 513~meV to 730~meV) with the elongation for the considered elongation range $\varepsilon = 9$--12\%.

\begin{figure}
\hfil\includegraphics{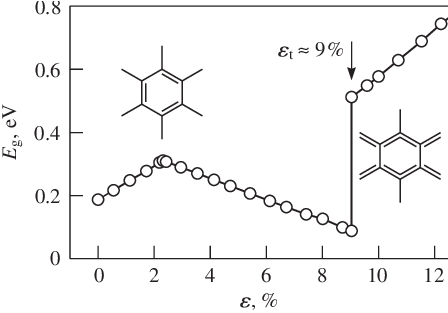}
\vspace{-6pt}
\caption{The band gap $E_g$ in the $\pi$-electron energy spectrum of the uniaxially deformed $(6,0)$ CNT versus its axial elongation $\varepsilon$. (The solid line is a guide to the eye).}\label{fig:05}
\end{figure}

\section{Strain sensor based on \emph{zigzag} CNT}

At present, the standard methods of the nanometer-scale detection of strain are based on micrometer size silicon piezoresistors or MOS transistors with piezoresistive channels~\cite{Lysko11}. Recently, a strain sensor based on commensurate\hspace{0pt}-\hspace{0pt}incommensurate phase transition in the double-walled CNTs was proposed~\cite{Bichoutskaia06}. Since the resistance of an undoped (intrinsic) semiconductor depends exponentially on the band gap, the predicted effect of the abrupt band gap change at the structural phase transition controlled by the axial elongation of the $(6,0)$ CNT can be detected by measurements of the CNT resistance. Thus, a strain sensor based on measurements of the resistance of the uniaxially stretched $(6,0)$ CNT can be proposed. The preliminary stretched $(6,0)$ CNT with the initial elongation $\varepsilon_{\rm ini}$ can be attached to electrodes which are positioned on controlled object (see Fig.~6). If this object is compressed or extended, the elongation of the CNT is $\varepsilon = \varepsilon_{\rm ini} + \varepsilon_{\rm obj}$, where $\varepsilon_{\rm obj}$ is the object elongation. At the object elongation $\varepsilon_{\rm obj} = \varepsilon_{\rm t} - \varepsilon_{\rm ini}$, which corresponds to the phase transition in the CNT, the abrupt change of the resistance can be detected. The proposed strain sensor can be used for detection both compression and extension of the object for the initial elongations of the CNT $\varepsilon_{\rm ini} > \varepsilon_{\rm t}$ and $\varepsilon_{\rm ini} < \varepsilon_{\rm t}$, respectively. 
Since the value of a typical diameter of a single-walled CNT is within several nm, a minimal size of a strain sensor is limited mainly by a minimal possible width of the electrodes.

The $(6,0)$ CNT is considered here as an example of $(3n,0)$ \emph{zigzag} CNTs. We believe that the phase transition from a narrow gap to a moderate gap semiconductor is possible also at elongation of other $(3n,0)$ CNTs. Thus, the proposed strain sensor can be based also on $(3n,0)$ CNTs as well as on $(n,n)$ armchair CNTs, where a narrow gap semiconductor to metal transition is predicted at elongation~\cite{Poklonski08CPL187}. A set of $(3n,0)$ or $(n,n)$ with different chiral indices (and thus with different critical elongations $\varepsilon_{\rm t}$) or with identical chiral indices but with different initial elongations $\varepsilon_{\rm ini}$ can be attached to the same pair of the electrodes. For such a scheme of the strain sensor, the transitions take place for different CNTs at different elongations of the controlled object, therefore the sensor can be used not only to detect the presence of strain in the controlled object but also to measure the value of this strain.

\begin{figure}
\hfil\includegraphics{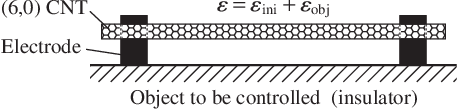}
\vspace{-6pt}
\caption{The nanometer-scale strain sensor based on the stretched $(6,0)$ CNT. If initial nanotube elongation is less than critical one $\varepsilon_{\rm ini} < \varepsilon_{\rm t}$, then the strain sensor operates for registering an extension of an object ($\varepsilon_{\rm obj} > 0$), and if $\varepsilon_{\rm ini} > \varepsilon_{\rm t}$, it operates for registering a compression of an object ($\varepsilon_{\rm obj} < 0$).}\label{fig:06}
\end{figure}

\section{Conclusions}

The first order deformational structural phase transition at zero temperature is revealed at the critical uniaxial elongation $\varepsilon_{\rm t} \approx 9\%$ of the $(6,0)$ carbon nanotube by semiempirical PM3 molecular orbital calculations. Namely, it is found that the $(6,0)$ nanotube has the Kekule structure with four types of bonds and the quinoid structure with two types of bonds for the elongations less and greater than the critical value $\varepsilon_{\rm t}$, respectively. Geometry optimization by the semiempirical PM3 method is found to be better than the DFT results as compared to experimental data on the C$_{60}$ molecule. Tight-binding band structure calculations show that this structural phase transition is followed by the transition in the electron subsystem from the narrow gap ($\approx 0.1$~eV) to the moderate gap ($\approx 0.5$~eV) semiconductor. A strain sensor of nanometer size based on measurements of resistance of the uniaxially stretched nanotube is proposed.

\section*{Acknowledgements}

This work has been partially supported by the BFBR (Grant Nos.~F11V-001, F12R-178) and RFBR (Grant 12-02-90041-Bel, 11-02-00604-a).
The authors thank Dr. D.B. Migas for discussions of the presented results.

\appendix
\setcounter{figure}{0}
\section{Tight-binding model details}
\subsection*{A.1. Kekule structure}
For the Kekule structure with four types of bond ($a$, $b$, $c$ and $d$) the translational unit cell (Fig.~1a) has three times more inequivalent atoms than the quinoid structure with two types of bonds $a'$ and $b'$ (Fig.~1b). Therefore, for the Kekule structure the tight-binding Hamiltonian matrix has dimensionality $6\times6$ which is three times greater than the one for the quinoid structure.

In the case of the Kekule structure with four different bond lengths (Fig.~1a) the two-dimensional primitive unit cell (Fig.~A.1) contains six carbon atoms (three $\alpha$ and three $\beta$ atoms). 

By employing the tight-binding approach for the Kekule structure, a Bloch wave function can be written as~\cite{Viet94, Saito98}:
\begin{equation}\label{eq:03}
   \Psi(\vec{r}) = \frac{1}{\sqrt{N_{\rm t}}}\sum_{\vec{R}_\zeta}\exp({\rm i}\vec{k}\vec{R}_\zeta)\,\psi_z(\vec{r} - \vec{R}_\zeta),
\end{equation}
where $N_{\rm t}$ is the number of translational unit cells, $\vec{R}_\zeta$ is the position of the $\zeta$-th kind of atoms ($\zeta = \alpha_1, \alpha_2, \alpha_3, \beta_1, \beta_2, \beta_3$), and $\psi_z(\vec{r})$ is the normalized $p_z$ atomic orbital wave function of an isolated carbon atom.

\begin{figure}
\hfil\includegraphics{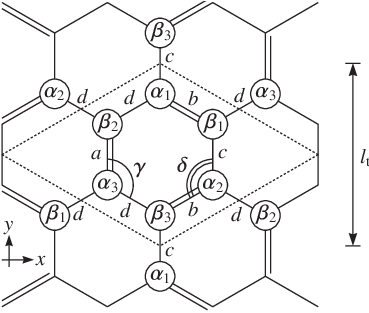}
\caption{The two-dimensional primitive unit cell (shown by dotted lines) of the \emph{zigzag} carbon nanotube with the Kekule structure. The coordinate system $x, y$ is used for calculations of the Hamiltonian matrix (\ref{eq:04}).}\label{fig:A1}
\end{figure}
In the tight-binding approximation, the calculations of a band structure of a CNT with the Kekule structure are reduced to diagonalization of the six order matrix. The Hamiltonian matrix is obtained as (see also~\cite{Poklonski09MS681}):
\begin{equation}\label{eq:04}
\text{\small\(\displaystyle
   \bordermatrix{ & \alpha_1 & \alpha_2 & \alpha_3 & \beta_1 & \beta_2 & \beta_3 \cr
\alpha_1&0&0&0&t_b{\rm e}^{-{\rm i}\vec{k}\vec{r}_1}&t_d{\rm e}^{-{\rm i}\vec{k}\vec{r}_2}&t_c{\rm e}^{-{\rm i}\vec{k}\vec{r}_3}\cr
\alpha_2&0&0&0&t_c{\rm e}^{-{\rm i}\vec{k}\vec{r}_3}&t_d{\rm e}^{-{\rm i}\vec{k}\vec{r}_4}&t_b{\rm e}^{-{\rm i}\vec{k}\vec{r}_5}\cr
\alpha_3&0&0&0&t_d{\rm e}^{-{\rm i}\vec{k}\vec{r}_2}&t_a{\rm e}^{-{\rm i}\vec{k}\vec{r}_6}&t_d{\rm e}^{-{\rm i}\vec{k}\vec{r}_4}\cr
\beta_1&t_b{\rm e}^{{\rm i}\vec{k}\vec{r}_1}&t_c{\rm e}^{{\rm i}\vec{k}\vec{r}_3}&t_d{\rm e}^{{\rm i}\vec{k}\vec{r}_2}&0&0&0\cr
\beta_2&t_d{\rm e}^{{\rm i}\vec{k}\vec{r}_2}&t_d{\rm e}^{{\rm i}\vec{k}\vec{r}_4}&t_a{\rm e}^{{\rm i}\vec{k}\vec{r}_6}&0&0&0\cr
\beta_3&t_c{\rm e}^{{\rm i}\vec{k}\vec{r}_3}&t_b{\rm e}^{{\rm i}\vec{k}\vec{r}_5}&t_d{\rm e}^{{\rm i}\vec{k}\vec{r}_4}&0&0&0},\hspace{-6pt}
\)}
\end{equation}
where $t_a, t_b, t_c, t_d$ are the hopping integrals corresponding to the bonds $a$, $b$, $c$, $d$, respectively; $\vec{k} = k_x\vec{e}_x + k_y\vec{e}_y$ is the electron wave-vector; $\vec{e}_x, \vec{e}_y$ are the unit vectors of the coordinate system; $\vec{r}_1 = \alpha_1\beta_1 = b_x\vec{e}_x - b_y\vec{e}_y$, $\vec{r}_2 = \alpha_1\beta_2 = -d_x\vec{e}_x - d_y\vec{e}_y$, $\vec{r}_3 = \alpha_1\beta_3 = c_y\vec{e}_y = c\vec{e}_y$, $\vec{r}_4 = \alpha_2\beta_2 = \alpha_3\beta_3 = d_x\vec{e}_x - d_y\vec{e}_y$, $\vec{r}_5 = \alpha_2\beta_3 = -b_x\vec{e}_x - b_y\vec{e}_y$, $\vec{r}_6 = \alpha_3\beta_2 = a_y\vec{e}_y = a\vec{e}_y$ are the vectors connecting nearest $\alpha_i$ and $\beta_i$ atoms; $a_x = 0$, $a_y = a$, $c_x = 0$, $c_y = c$, $b_x = b\sin\delta$, $b_y = -b\cos\delta$, $d_x = d\sin\gamma$, $d_y = -d\cos\gamma$ are the projections of the bonds on the $x$ and $y$ axes of the Cartesian coordinate system; $\gamma$ and $\delta$ are the angles between the bonds $a$ and $d$, and the bonds $b$ and $c$, respectively.

The periodic boundary condition is
\[
   \vec{C}_{\rm h}{\cdot}\vec{k} = 2\pi q;\quad
   q = 1, \ldots, 2n/3,
\]
\[
   k_x = \frac{2\pi q}{C_{\rm h}} = \frac{\pi q}{2(2d_x + b_x)},\quad
   -\frac{\pi}{l_{\rm t}} \leq k_y \leq \frac{\pi}{l_{\rm t}}, 
\]
where $C_{\rm h} = 4(2d_x + b_x)$ is the magnitude of the chiral vector $\vec{C}_{\rm h} = n\vec{a}_1 + 0{\cdot}\vec{a}_2 \equiv (n,0)$ (for $n = 6$ the value of $C_{\rm h}$ is a circumference length of the $(6,0)$ CNT; $\vec{a}_1$ and $\vec{a}_2$ are the unit vectors of graphene), and $l_{\rm t} = a + c + 2d_y$ is the period of the $(6,0)$ CNT with the Kekule structure along the nanotube axis.

\begin{figure}
\hfil\includegraphics{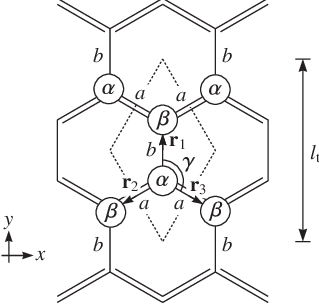}
\caption{The two-dimensional primitive unit cell (shown by dotted lines) of the \emph{zigzag} carbon nanotube with the quinoid structure. The coordinate system $x, y$ is used for calculations of the Hamiltonian matrix (\ref{eq:05}).}\label{fig:A2}
\end{figure}

\subsection*{A.2. Quinoid structure}
In the case of a CNT with the quinoid structure with two different bond lengths ($a'$ and $b'$ as shown in Fig.~1b) the primitive unit cell contains two atoms (denoted as $\alpha$ and $\beta$ in Fig.~A.2). In this section we will denote bonds $a'$ and $b'$ as $a$ and $b$, respectively.

The Bloch wave function for the quinoid structure can be written in form of Eq.~(\ref{eq:03}), where $\vec{R}_\zeta$ stands for the positions $\vec{R}_\alpha$ and $\vec{R}_\beta$ of the $\alpha$ and $\beta$ carbon atoms.

The tight-binding Hamiltonian matrix in this case has the form~\cite{Okahara94, Saito98}
\begin{equation}\label{eq:05}
   \arraycolsep=1pt
   H = \left[\begin{matrix}0&H_{\alpha\beta}^*\\H_{\alpha\beta}&0\end{matrix}\right],
\end{equation}
where $H_{\alpha\beta} = t_b{\rm e}^{{\rm i}\vec{k}\vec{r}_1} + t_a{\rm e}^{{\rm i}\vec{k}\vec{r}_2} + t_a{\rm e}^{{\rm i}\vec{k}\vec{r}_3}$ and $H_{\alpha\beta}^* = t_b{\rm e}^{-{\rm i}\vec{k}\vec{r}_1} + t_a{\rm e}^{-{\rm i}\vec{k}\vec{r}_2} + t_a{\rm e}^{-{\rm i}\vec{k}\vec{r}_3}$.

As it follows from Fig.~A.2 the vectors $\vec{r}_1$, $\vec{r}_2$, $\vec{r}_3$ are defined in the chosen reference system as:
\[
   \vec{r}_1 = b\vec{e}_y,\quad
   \vec{r}_2 = -a_x\vec{e}_x - a_y\vec{e}_y,\quad
   \vec{r}_3 = a_x\vec{e}_x - a_y\vec{e}_y,
\]
where $a_x = a\sin\gamma$, $a_y = -a\cos\gamma$ are the projections of the bond $a$ on the $x$ and $y$ axes; the wave-vector is $\vec{k} = k_x\vec{e}_x + k_y\vec{e}_y$.

The periodic boundary condition is
\[
   \vec{C}_{\rm h}{\cdot}\vec{k} = 2\pi q;\quad
   q = 1, \ldots, 2n,
\]
\[
   k_x = \frac{2\pi q}{C_{\rm h}} = \frac{\pi q}{6a_x},\quad
   -\frac{\pi}{l_{\rm t}} \leq k_y \leq \frac{\pi}{l_{\rm t}},
\]
where $C_{\rm h} = 2na_x = 12a_x$ is the magnitude of the chiral vector, and $l_{\rm t} = 2(a_y + b)$ is the period of the $(6,0)$ CNT with the quinoid structure along the nanotube axis.

To obtain the energy spectrum the following secular equation should be solved 
\[
   \arraycolsep=4pt
   \det\left[\begin{matrix}-E&H_{\alpha\beta}^*\\H_{\alpha\beta}&-E\end{matrix}\right] = 0.
\]
By expanding the above determinant, we obtain
\[
   E^2 = t_b^2 + 4t_a^2\cos^2(\pi q/6) + 4t_at_b\cos(k_yl_{\rm t}/2)\cos(\pi q/6).
\]

Thus, in the tight-binding approximation the dependence of $\pi$-electron energy on the wave vector along the nanotube axis $E_m(k)$ (band structure) for a $(n,0)$ CNT with the quinoid structure is~\cite{Okahara94}:
\begin{align}\label{eq:06}
   E_m(k) = &\pm[t_b^2 + 4t_a^2\cos^2(\pi q/n) \notag\\
      &+ 4t_at_b\cos(\pi q/n)\cos(kl_{\rm t}/2)]^{1/2},
\end{align}
where $t_a$ and $t_b$ are the hopping integrals for the bonds $a$ and $b$, respectively, $q = 1, 2, \ldots, 2n$ is an integer, and $-\pi/l_{\rm t} \leq k_y \leq \pi/l_{\rm t}$ is the wavenumber of the $\pi$-electron along the nanotube axis $y$.

For the CNT with the quinoid structure there are $4n$ energy bands in the electronic spectrum of the CNT, $4n - 4$ of which are twofold degenerate, according to equivalence of two possible directions of the electron quasi-momentum. Only bands with indexes $q$ equal to $n$ and $2n$ are not degenerate. The signs plus and minus in the formula (\ref{eq:06}) correspond to the conduction and valence bands, respectively, which are symmetric relative to the Fermi level $E = 0$. For the $(n,0)$ CNT with even $n$, bands with index $q = n/2$ are dispersionless (the electron energy does not depend on the quasi-momentum) (see also~\cite{Saito98}). The formula (\ref{eq:06}) for the electron energy bands shows that the electron energy in the $(n,0)$ CNT can be equal to zero only when $n$ is multiple of 3 and $t_a = t_b$ (i.e., all equal C--C bonds). Only in this case the electric conductivity of the CNT has a metallic character, in all other cases there is a gap in the electron energy spectrum.



\bibliographystyle{elsarticle-num}
\bibliography{poklonski}







\end{document}